\documentclass[12pt,sort&compress]{elsarticle}
\usepackage[T1]{fontenc}
\usepackage[latin9]{inputenc}
\usepackage{amsmath}
\usepackage{amssymb}
\usepackage{graphicx}

\makeatletter

\providecommand{\tabularnewline}{\\}

\journal{Physica B: Condensed Matter}

\makeatother

\begin{document}

\begin{frontmatter}{}

\title{Single-particle properties of the Hubbard model\\
in a novel three-pole approximation}

\author[uni]{Andrea Di Ciolo}

\author[uni,cnrsa,cnism]{Adolfo Avella}

\address[uni]{Dipartimento di Fisica ``E.R. Caianiello'', Universit\`a degli
Studi di Salerno, I-84084 Fisciano (SA), Italy}

\address[cnrsa]{CNR-SPIN, UOS di Salerno, I-84084 Fisciano (SA), Italy}

\address[cnism]{Unit\`a CNISM di Salerno, Universit\`a degli Studi di Salerno,
I-84084 Fisciano (SA), Italy}
\begin{abstract}
We study the 2D Hubbard model using the Composite Operator Method
within a novel three-pole approximation. Motivated by the long-standing
experimental puzzle of the single-particle properties of the underdoped
cuprates, we include in the operatorial basis, together with the usual
Hubbard operators, a field describing the electronic transitions dressed
by the nearest-neighbor spin fluctuations, which play a crucial role
in the unconventional behavior of the Fermi surface and of the electronic
dispersion. Then, we adopt this approximation to study the single-particle
properties in the strong coupling regime and find an unexpected behavior
of the van Hove singularity that can be seen as a precursor of a pseudogap
regime.
\end{abstract}

\end{frontmatter}{}

\section{Introduction}

The Hubbard model \citep{Hubbard_63,Hubbard_64,Hubbard_64a}, and
its derivatives \citep{Chao_77,Chao_78} and extensions \citep{Emery_87,Emery_88,Emery_94,Avella_13b},
constitutes still one of the most studied model in condensed matter
theory because of its relevance to almost all strongly correlated
systems and, in particular, transition metal oxides. We report on
a solution of the two-dimensional Hubbard model in the framework of
the Composite Operator Method (COM) \citep{Mancini_04,Avella_12a,Avella_14}
within a novel three-pole approximation \citep{Avella_14}. The COM,
which is based on the equations of motion and Green's function formalisms,
is highly tunable and expressly devised for the characterization of
strongly correlated electronic states and the exploration of novel
emergent phases. Motivated by the long-standing experimental challenge
posed by the puzzling single-particle properties of the underdoped
cuprates \citep{Timusk_99,Avella_14a} (Fermi arcs, pseudogap, non-Fermi
liquid behavior, extreme momentum dependence of spectral properties,
\dots ) , we adopt a basis of fields containing the two Hubbard operators
and a third operator specifically designed to describe the electronic
transitions dressed by the nearest-neighbor spin fluctuations and
to capture the effects of these latter on all electronic properties.
The spin fluctuations may play a crucial role in the mechanism of
pseudogap formation and evolution as well as in the unconventional
behavior of the Fermi surface, the spectral weights and the electronic
dispersion \citep{Bulut_94,Sadovskii_05,Avella_07}. Thus, we have
designed this non canonical, but very efficient, operatorial representation
of correlated electrons to provide a reliable analytical computational
tool. The quality of this approximation has already been assessed
positively by comparing its results against the numerical ones for
many integrated/local quantities and for the band dispersion \citep{Avella_14}.
In this short paper, we adopt this approximation to study the single-particle
properties of the model in the strong coupling regime, where the effects
of the spin fluctuations, accurately treated in our approach, are
more relevant and can induce unconventional features in all analyzed
spectral properties. In particular, we find an unexpected behavior
of the van Hove singularity only for high enough values of the on-site
Coulomb repulsion that can be seen as a precursor of a pseudogap regime.

\section{Model and method}

The two-dimensional Hubbard model reads as
\begin{align}
H & =\sum_{\mathbf{i}}\left(-4tc^{\dagger}\left(i\right)\cdot c^{\alpha}\left(i\right)+Un_{\uparrow}\left(i\right)n_{\downarrow}\left(i\right)-\mu n\left(i\right)\right)\label{eq:Ham}
\end{align}
 where $c^{\dagger}\left(i\right)=\begin{pmatrix}c_{\uparrow}^{\dagger}\left(i\right) & c_{\downarrow}^{\dagger}\left(i\right)\end{pmatrix}$
is the electronic field operator, in spinorial notation ($\cdot$
stands for the inner (scalar) product in spin space) and Heisenberg
picture ($i=\left(\mathbf{i},t_{i}\right)$, being $\mathbf{i}=\mathbf{r_{i}}$
a Bravais lattice vector and $t_{i}$ the time), and $\sigma=\uparrow,\downarrow$
the electronic spin. $n_{\sigma}\left(i\right)=c_{\sigma}^{\dagger}\left(i\right)c_{\sigma}\left(i\right)$
is the particle density operator for spin $\sigma$ at site $\mathbf{i}$
and $n\left(i\right)=\sum_{\sigma}n_{\sigma}\left(i\right)=c^{\dagger}\left(i\right)\cdot c\left(i\right)$
is the total particle density operator at site $\mathbf{i}$. $c^{\alpha}\left(\mathbf{i},t\right)=\sum_{\mathbf{j}}\alpha_{\mathbf{ij}}c\left(\mathbf{j},t\right)$
where $\alpha_{\mathbf{ij}}=\dfrac{1}{2d}\delta_{\left\langle \mathbf{i}\mathbf{j}\right\rangle }$
is the nearest-neighbor projector. $U$ is the on-site Coulomb repulsion,
$t$ is the nearest-neighbor hopping integral and $\mu$ is the chemical
potential.

According the the COM recipe \citep{Avella_12a}, we adopt a three-field
basis

\begin{equation}
\Psi\left(i\right)=\left(\begin{array}{c}
\psi_{1}\left(i\right)\\
\psi_{2}\left(i\right)\\
\psi_{3}\left(i\right)
\end{array}\right)=\left(\begin{array}{l}
\xi\left(i\right)=\left(1-n\left(i\right)\right)c\left(i\right)\\
\eta\left(i\right)=n\left(i\right)c\left(i\right)\\
c_{s}\left(i\right)=n_{k}\left(i\right)\sigma_{k}\cdot c^{\alpha}\left(i\right)
\end{array}\right)\label{eq:Psi3p}
\end{equation}
where $\eta\left(i\right)$ and $\xi\left(i\right)$ are the local
Hubbard operators describing the transitions which change the electron
numbers per site from 2 to 1 and from 1 to 0 respectively, and $c_{s}\left(i\right)$
describes the electronic transitions dressed by nearest-neighbor spin
fluctuations, being $n_{k}\left(i\right)=c^{\dagger}\left(i\right)\cdot\sigma_{k}\cdot c\left(i\right)$
the spin density operator and $\sigma_{k}$ the Pauli matrices. We
choose $c_{s}$ as third basis component according to the idea that
the spin fluctuations are a key ingredient to describe strong correlations
and they influence the dynamics more substantially than the other
types of fluctuations (charge, pair, ...). The operators $\psi_{n}$
are products of $c$ operators and, accordingly, they are composite
operators. The vectorial current $J$ of the basis, $J\left(i\right)=i\dfrac{\partial}{\partial t}\Psi\left(i\right)=[\Psi\left(i\right),H]$,
can be rewritten as $J\left(i\right)=\sum_{\mathbf{j}}\varepsilon\left(\mathbf{i},\mathbf{j}\right)\Psi\left(\mathbf{j},t\right)+\delta J\left(i\right)$,
where the first term represents the operatorial projection of the
current $J$ on the basis $\Psi$ and the second term is the residual
current $\delta J$. The energy matrix $\varepsilon$ is obtained
by means of the constraint $\left\langle \left\{ \delta J\left(\mathbf{i},t\right),\Psi^{\dagger}\left(\mathbf{j},t\right)\right\} \right\rangle =0$,
which assures that $\delta J$ retains only the physics orthogonal
to the relevant one described by the chosen basis $\Psi$: $\varepsilon\left(\mathbf{k}\right)=m\left(\mathbf{k}\right)I^{-1}\left(\mathbf{k}\right)$.
We have introduced, for the sake of simplicity, the $m$-matrix $m\left(\mathbf{i},\mathbf{j}\right)=\left\langle \left\{ J\left(\mathbf{i},t\right),\Psi^{\dagger}\left(\mathbf{j},t\right)\right\} \right\rangle =\frac{1}{N}\sum_{\mathbf{k}}e^{i\mathbf{k\cdot\left(r_{i}-r_{j}\right)}}m\left(\mathbf{k}\right)$
and the normalization matrix $I\left(\mathbf{i},\mathbf{j}\right)=\left\langle \left\{ \Psi\left(\mathbf{i},t\right),\:\Psi^{\dagger}\left(\mathbf{j},t\right)\right\} \right\rangle =\frac{1}{N}\sum_{\mathbf{k}}e^{i\mathbf{k\cdot\left(r_{i}-r_{j}\right)}}I\left(\mathbf{k}\right)$;
$\left\langle ...\right\rangle $ stands for the thermal average in
the grand-canonical ensemble and $\mathbf{k}$ runs over the first
Brillouin zone. $\varepsilon\left(\mathbf{k}\right)$, which is the
Fourier transform of $\varepsilon\left(\mathbf{i},\mathbf{j}\right)$
\textendash{} $\varepsilon\left(\mathbf{i},\mathbf{j}\right)=\frac{1}{N}\sum_{\mathbf{k}}e^{i\mathbf{k\cdot\left(r_{i}-r_{j}\right)}}\varepsilon\left(\mathbf{k}\right)$
\textendash , has real eigenvalues $E^{\left(\nu\right)}\left(\mathbf{k}\right)$,
which represent the excitation energy spectrum of the system, and
its eigenvectors identify the elementary excitations of the system,
within this approximation. We consider the thermal retarded Green's
function (GF) $G(\emph{i},\emph{j})=\langle\mathcal{R}\left[\Psi(\emph{i})\Psi^{\dagger}(\emph{j})\right]\rangle$
and its Fourier transform $G\left(\mathbf{k,\omega}\right)$, which
can be obtained solving its Dyson's equation in the frequency-momentum
space (once the residual current is neglected)
\begin{equation}
G(\boldsymbol{k},\:\omega)=\dfrac{1}{\omega-\epsilon(\boldsymbol{k})+\mathrm{i}\delta}I(\boldsymbol{k})=\sum_{\nu}\dfrac{\sigma^{(\nu)}(\boldsymbol{k})}{\omega-E^{\left(\nu\right)}(\boldsymbol{k})+\mathrm{i}\delta}\label{eq:Gmnpole}
\end{equation}
$E^{\left(\nu\right)}(\boldsymbol{k})$ act as bands of the system
and $\sigma^{\left(\nu\right)}\left(\mathbf{k}\right)$ are the matricial
spectral density weights per band $\sigma_{mn}^{(\nu)}(\boldsymbol{k})=\sum_{c}\Omega_{m\nu}(\boldsymbol{k})\Omega_{\nu c}^{-1}(\boldsymbol{k})I_{cn}(\boldsymbol{k})$,
where the matrix $\Omega\left(\mathbf{k}\right)$ has the eigenvectors
of $\varepsilon\left(\mathbf{k}\right)$ as columns. The correlation
functions of the fields of the basis $\Psi$, $C_{mn}\left(\mathbf{i},\mathbf{j}\right)=\left\langle \psi_{m}\left(\mathbf{i}\right)\psi_{n}^{\dagger}\left(\mathbf{j}\right)\right\rangle $,
can be determined in terms of the GF by means of the spectral theorem
\begin{equation}
C_{mn}\left(\mathbf{k},\omega\right)=2\pi\sum\limits _{\nu}\left[1-f_{\mathrm{F}}\left(E^{\left(\nu\right)}\left(\mathbf{k}\right)\right)\right]\sigma_{mn}^{\left(\nu\right)}\left(\mathbf{k}\right)\delta\left(\omega-E^{\left(\nu\right)}\left(\mathbf{k}\right)\right)\label{eq:Ck}
\end{equation}
where $f_{\mathrm{F}}$ is the Fermi function.

\subsection{The equations of motion}

The fields $\xi(i)$ and $\eta(i)$ satisfy the following equations
of motion 
\begin{align}
 & \mathrm{i}\frac{\partial}{\partial t}\xi\left(i\right)=-\mu\xi\left(i\right)-4tc^{\alpha}\left(i\right)-4t\pi\left(i\right)\label{eq:em-xi}\\
 & \mathrm{i}\frac{\partial}{\partial t}\eta\left(i\right)=\left(U-\mu\right)\eta\left(i\right)+4t\pi\left(i\right)\label{eq:em-eta}
\end{align}
where $\pi\left(i\right)=\frac{1}{2}n_{\mu}(i)\sigma^{\mu}\cdot c^{\alpha}\left(i\right)+c^{\dagger\alpha}\left(i\right)\cdot c\left(i\right)\otimes c\left(i\right)$
is a higher-order composite field, $n_{\mu}\left(i\right)=c^{\dagger}\left(i\right)\cdot\sigma_{\mu}\cdot c\left(i\right)$
is the charge- ($\mu=0$) and spin- ($\mu=1,2,3=k$) density operator,
$\sigma_{0}=\mathbf{1}$ is the identity matrix and $\otimes$ stands
for the outer product in spin space. The field $c_{s}\left(i\right)$
satisfies the following equation of motion
\begin{equation}
\mathrm{i}\frac{\partial}{\partial t}c_{s}\left(i\right)=-\mu c_{s}\left(i\right)+4t\kappa_{s}\left(i\right)+U\eta_{s}\left(i\right)\label{eq:em-cs}
\end{equation}
where $\kappa_{s}\left(i\right)=\left(c^{\alpha\dagger}\left(i\right)\cdot\sigma_{k}\cdot c\left(i\right)-c^{\dagger}\left(i\right)\cdot\sigma_{k}\cdot c^{\alpha}\left(i\right)\right)\sigma_{k}\cdot c^{\alpha}\left(i\right)-n_{k}\left(i\right)\sigma_{k}\cdot c^{\alpha^{2}}\left(i\right)$,
$\eta_{s}\left(i\right)=n_{k}\left(i\right)\sigma_{k}\cdot\eta^{\alpha}\left(i\right)$
and $c^{\alpha^{2}}\left(\mathbf{i},t\right)=\sum_{\mathbf{j}\mathbf{l}}\alpha_{\mathbf{ij}}\alpha_{\mathbf{j}\mathbf{l}}c\left(\mathbf{l},t\right)$.

\subsection{The normalization matrix $I$\label{sec:I}}

The normalization $I\left(\mathbf{k}\right)$ matrix is symmetric
by construction and its entries have the following expressions
\begin{align}
 & I_{11}\left(\mathbf{k}\right)=I_{11}=1-\frac{n}{2},\quad I_{12}\left(\mathbf{k}\right)=0,\quad I_{22}\left(\mathbf{k}\right)=I_{22}=\frac{n}{2}\label{eq:I11k}\\
 & I_{13}\left(\mathbf{k}\right)=3C_{\xi c}^{\alpha}+\frac{3}{2}\alpha\left(\mathbf{k}\right)\chi_{s}^{\alpha}\label{eq:I13k}\\
 & I_{23}\left(\mathbf{k}\right)=3C_{\eta c}^{\alpha}-\frac{3}{2}\alpha\left(\mathbf{k}\right)\chi_{s}^{\alpha}\label{eq:I23k}\\
 & I_{33}\left(\mathbf{k}\right)\cong4C_{c_{s}c}^{\alpha}+\frac{3}{2}C_{\eta\eta}+3\alpha\left(\mathbf{k}\right)\left(f_{s}+\frac{1}{4}C_{cc}^{\alpha}\right)
\end{align}
where $n=\left\langle n\left(i\right)\right\rangle $ is the filling,
$\chi_{s}^{\alpha}=\frac{1}{3}\sum_{k}\left\langle n_{k}^{\alpha}\left(i\right)n_{k}\left(i\right)\right\rangle $
is the nearest-neighbor spin-spin correlation function, $f_{s}=\frac{1}{3}\left\langle c^{\dagger}\left(i\right)\cdot\sigma_{k}\cdot c^{\alpha}\left(i\right)n_{k}^{\alpha}\left(i\right)\right\rangle $
is a higher-order (up to three different sites are involved) spin-spin
correlation function, and $C_{mn}^{\alpha}=\left\langle \psi_{m}^{\alpha}\left(\mathbf{i}\right)\psi_{n}^{\dagger}\left(\mathbf{i}\right)\right\rangle $.
Higher-order terms (involving more distant sites) in $I_{33}\left(\mathbf{k}\right)$
have been neglected \citep{Avella_14}.

\subsection{The $m$-matrix\label{sec:m}}

The $m\left(\mathbf{k}\right)$ matrix is symmetric by construction
and its entries have the following expressions
\begin{align}
 & m_{11}\left(\mathbf{k}\right)=-\mu I_{11}-4t\left[\Delta+\left(p+I_{11}-I_{22}\right)\alpha\left(\mathbf{k}\right)\right]\label{eq:m11k}\\
 & m_{12}\left(\mathbf{k}\right)=4t\left[\Delta+\left(p-I_{22}\right)\alpha\left(\mathbf{k}\right)\right]\label{eq:m12k}\\
 & m_{13}\left(\mathbf{k}\right)=-\left(\mu+4t\alpha\left(\mathbf{k}\right)\right)I_{13}\left(\mathbf{k}\right)-4t\alpha\left(\mathbf{k}\right)I_{23}\left(\mathbf{k}\right)-2tI_{33}\left(\mathbf{k}\right)-4t\gamma_{m}\alpha\left(\mathbf{k}\right)\\
 & m_{22}\left(\mathbf{k}\right)=\left(U-\mu\right)I_{22}-4t\left[\Delta+p\alpha\left(\mathbf{k}\right)\right]\label{eq:m22k}\\
 & m_{23}\left(\mathbf{k}\right)=\left(U-\mu\right)I_{23}\left(\mathbf{k}\right)+2tI_{33}\left(\mathbf{k}\right)+4t\gamma_{m}\alpha\left(\mathbf{k}\right)\label{eq:m23k}\\
 & m_{33}\left(\mathbf{k}\right)\cong-\mu I_{33}\left(\mathbf{k}\right)+m_{33}^{0r}+m_{33}^{\alpha r}\alpha\left(\mathbf{k}\right)\label{eq:m33k}
\end{align}
where $\Delta=C_{\xi\xi}^{\alpha}-C_{\eta\eta}^{\alpha}$, $p=\frac{1}{4}\left(\chi_{0}^{\alpha}+3\chi_{s}^{\alpha}\right)-\chi_{p}^{\alpha}$
is a combination of the nearest-neighbor charge-charge $\chi_{0}^{\alpha}=\left\langle n^{\alpha}\left(i\right)n\left(i\right)\right\rangle $,
spin-spin $\chi_{s}^{\alpha}$ and pair-pair $\chi_{p}^{\alpha}=\left\langle \left[c_{\uparrow}\left(i\right)c_{\downarrow}\left(i\right)\right]^{\alpha}c_{\downarrow}^{\dagger}\left(i\right)c_{\uparrow}^{\dagger}\left(i\right)\right\rangle $
correlation functions, $\gamma_{m}$, $m_{33}^{0r}$, and $m_{33}^{\alpha r}$
are the combinations of many higher-order correlation functions \citep{Avella_14}.
Higher-order terms (involving more distant sites) in $m_{33}\left(\mathbf{k}\right)$
have been neglected \citep{Avella_14}.

\subsection{Self-consistency and Algebra constraints\label{sec:self}}

Algebra Constraints (ACs), exact relationship between the correlation
functions of the fields of the chosen operatorial basis dictated by
the non-canonical algebra they close, offer a very reliable way to
fix unknown parameters and allow, at the same time, to impose to the
system under analysis algebraic relations and/or symmetry requirements
that are valid for any coupling and any value of the external parameters.
In this case, we can recognize the following exact Algebra Constraints
\begin{align}
& C_{\xi\xi}=1-n+D,\quad C_{\xi\eta}=0,\quad C_{\eta\eta}=\frac{n}{2}-D\label{eq:Cxixi}\\
& C_{\xi c_{s}}=3C_{\xi c}^{\alpha},\quad C_{\eta c_{s}}=0\label{eq:Cxics}
\end{align}
where $D=\left\langle n_{\uparrow}\left(i\right)n_{\downarrow}\left(i\right)\right\rangle $
is the double occupancy. These relations lead to the following very
relevant ones: $n=2\left(1-C_{\xi\xi}-C_{\eta\eta}\right)$ and $D=1-C_{\xi\xi}-2C_{\eta\eta}$.
On the other hand, we can compute $\chi_{0}^{\alpha}$, $\chi_{s}^{\alpha}$,
$\chi_{p}^{\alpha}$ and $f_{s}$ by operatorial projection \citep{Avella_14}
\begin{align}
& \chi_{0}^{\alpha}\approx n^{2}-2\frac{I_{11}\left(C_{c\eta}^{\alpha}\right)^{2}+I_{22}\left(C_{c\xi}^{\alpha}\right)^{2}}{C_{\eta\eta}} \label{eq:chia0}\\
& \chi_{s}^{\alpha}\approx-2\frac{I_{11}\left(C_{c\eta}^{\alpha}\right)^{2}+I_{22}\left(C_{c\xi}^{\alpha}\right)^{2}}{2I_{11}I_{22}-C_{\eta\eta}}\\
 & \chi_{p}^{\alpha}\approx\frac{C_{c\xi}^{\alpha}C_{\eta c}^{\alpha}}{C_{\eta\eta}}\\
& f_{s}\approx-\frac{1}{2}C_{c\xi}^{\alpha}-\frac{3}{4}\chi_{s}^{\alpha}\left(\frac{C_{c\xi}^{\alpha}}{I_{11}}-\frac{C_{c\eta}^{\alpha}}{I_{22}}\right)\nonumber \\
& \quad-2\frac{C_{c\xi}^{\alpha}}{I_{11}}\left(C_{c\xi}^{\alpha^{2}}-\frac{1}{4}C_{c\xi}\right)-2\frac{C_{c\eta}^{\alpha}}{I_{22}}\left(C_{c\eta}^{\alpha^{2}}-\frac{1}{4}C_{c\eta}\right)
\end{align}
and use the three left ACs to compute $\gamma_{m}$, $m_{33}^{0r}$,
and $m_{33}^{\alpha r}$.

\section{Single-particle properties}

We can now analyze the behavior of the single-particle properties
of the system: energy bands, density of states and Fermi surface.
In particular, we can study the energy bands of the system $E^{\left(\nu\right)}\left(\mathbf{k}\right)$
along the principal directions of the first Brillouin zone ($\Gamma=\left(0,0\right)\rightarrow S=\left(\pi/2,\pi/2\right)\rightarrow M=\left(\pi,\pi\right)\rightarrow X=\left(\pi,0\right)\rightarrow\Gamma=\left(0,0\right)$)
as well as the corresponding electronic spectral density weight $\sigma_{cc}^{\left(\nu\right)}\left(\mathbf{k}\right)=\sum_{n,m=1}^{2}\sigma_{nm}^{\left(\nu\right)}\left(\mathbf{k}\right)$.
This latter corresponds to the component per band of the momentum
distribution function per spin $n\left(\mathbf{k}\right)$ at $T=0$
for those bands and momenta below the chemical potential. Giving to
each energy band $E^{\left(\nu\right)}\left(\mathbf{k}\right)$ a
thickness proportional to $\sigma_{cc}^{\left(\nu\right)}\left(\mathbf{k}\right)$
shows the effective relevance of each energy band, momentum per momentum,
with respect to actual occupation and possible hole/electron doping.
The density of states, $N\left(\omega\right)=\frac{1}{N}\sum_{\mathbf{k}}\sum_{\nu}\sigma_{cc}^{\left(\nu\right)}\left(\mathbf{k}\right)\delta\left(\omega-E^{\left(\nu\right)}\left(\mathbf{k}\right)\right)$,
depends on both the electronic spectral weight, $\sigma_{cc}^{\left(\nu\right)}\left(\mathbf{k}\right)$,
and the actual \emph{shape} (through the curvature $\nabla_{\mathbf{k}}E^{\left(\nu\right)}\left(\mathbf{k}\right)$)
of the bands
\begin{equation}
\delta\left(\omega-E^{\left(\nu\right)}\left(\mathbf{k}\right)\right)=\sum_{p}\frac{\delta\left(\mathbf{k}-\mathbf{k}_{p}^{\left(\nu\right)}\left(\omega\right)\right)}{\left|\nabla_{\mathbf{k}}E^{\left(\nu\right)}\left(\mathbf{k}\right)\right|_{\mathbf{k}=\mathbf{k}_{p}^{\left(\nu\right)}}}\label{eq:deltaEk-1}
\end{equation}
where $\mathbf{k}_{p}^{\left(\nu\right)}\left(\omega\right)$ are
the zeros of $\omega-E^{\left(\nu\right)}\left(\mathbf{k}\right)=0$.
Finally, we can investigate the shape of the Fermi surface by means
of the spectral function $A\left(\mathbf{k},\omega\right)=-\dfrac{1}{\pi}\Im\left[G_{cc}\left(\mathbf{k},\omega\right)\right]=\sum_{\nu}\sigma_{cc}^{\left(\nu\right)}\left(\mathbf{k}\right)\delta\left(\omega-E^{\left(\nu\right)}\left(\mathbf{k}\right)\right)$,
where $G_{cc}\left(\mathbf{k}\right)=\sum_{n,m=1}^{2}G_{nm}\left(\mathbf{k}\right)$
is the electronic Green's function. In fact, the position of the maxima
of $A\left(\mathbf{k},\omega=0\right)$ provides the effective Fermi
Surface as measured by ARPES (Angle Resolved Photo-Emission Spectroscopy)
experiments. In all calculations, $\delta\left(\omega\right)$ has
been replaced by a Lorentzian function $\dfrac{1}{\pi}\dfrac{\varepsilon}{\omega^{2}+\varepsilon^{2}}$
with $\varepsilon=0.05$.

\begin{figure}[!t]
\noindent \begin{centering}
\begin{tabular}{cc}
\includegraphics[height=4.5cm]{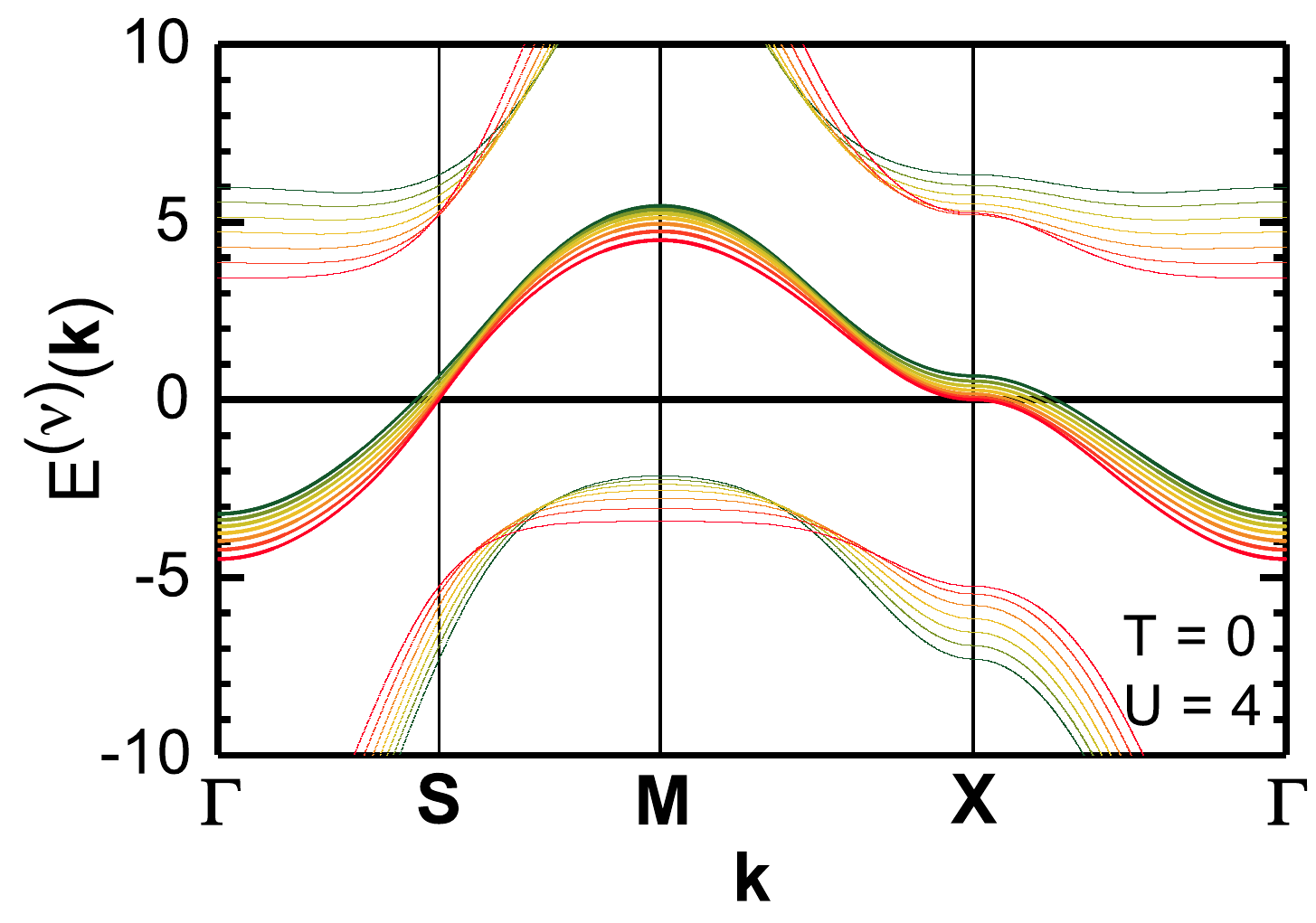} & \includegraphics[height=4.5cm]{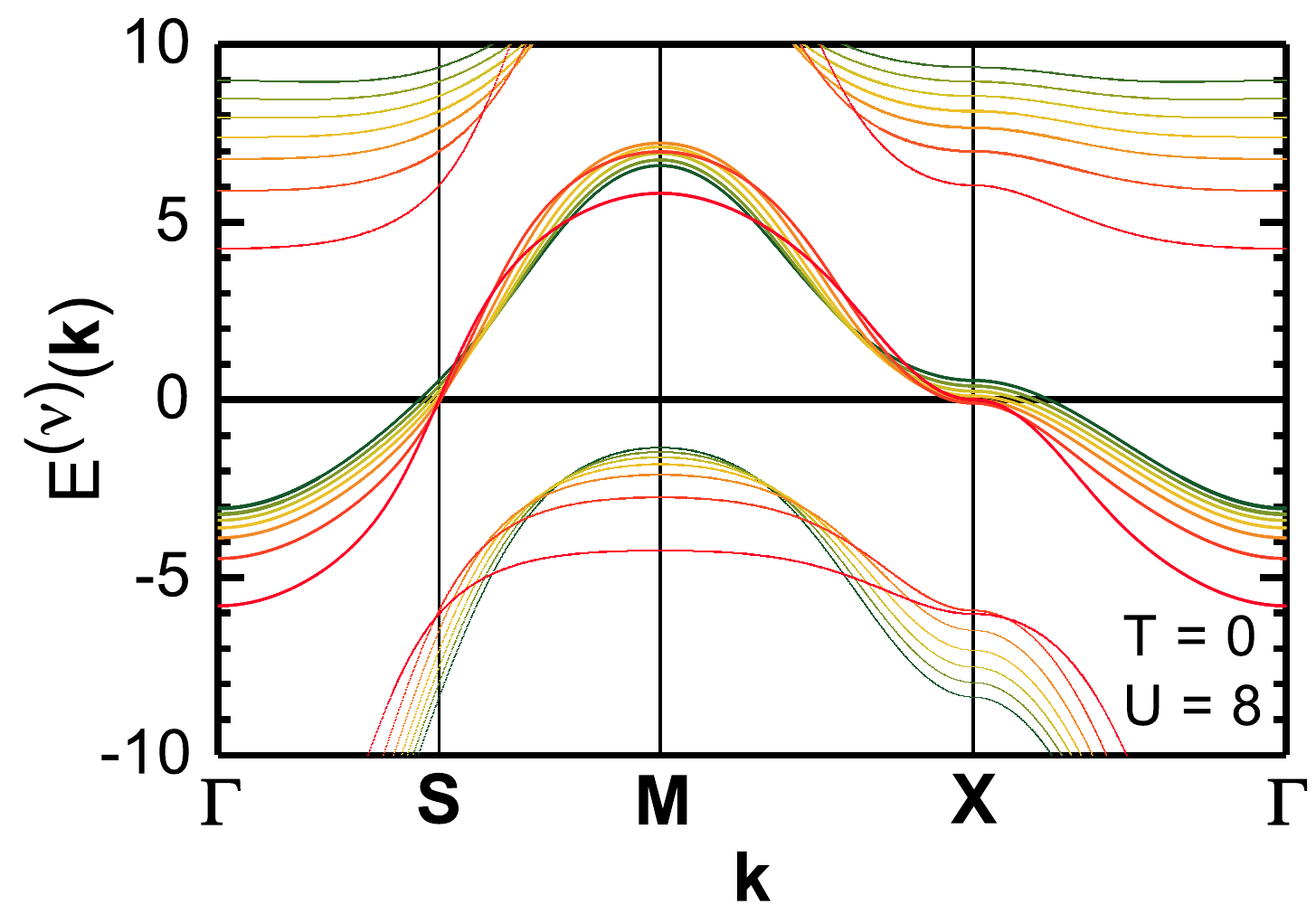}\tabularnewline
\includegraphics[height=5.5cm]{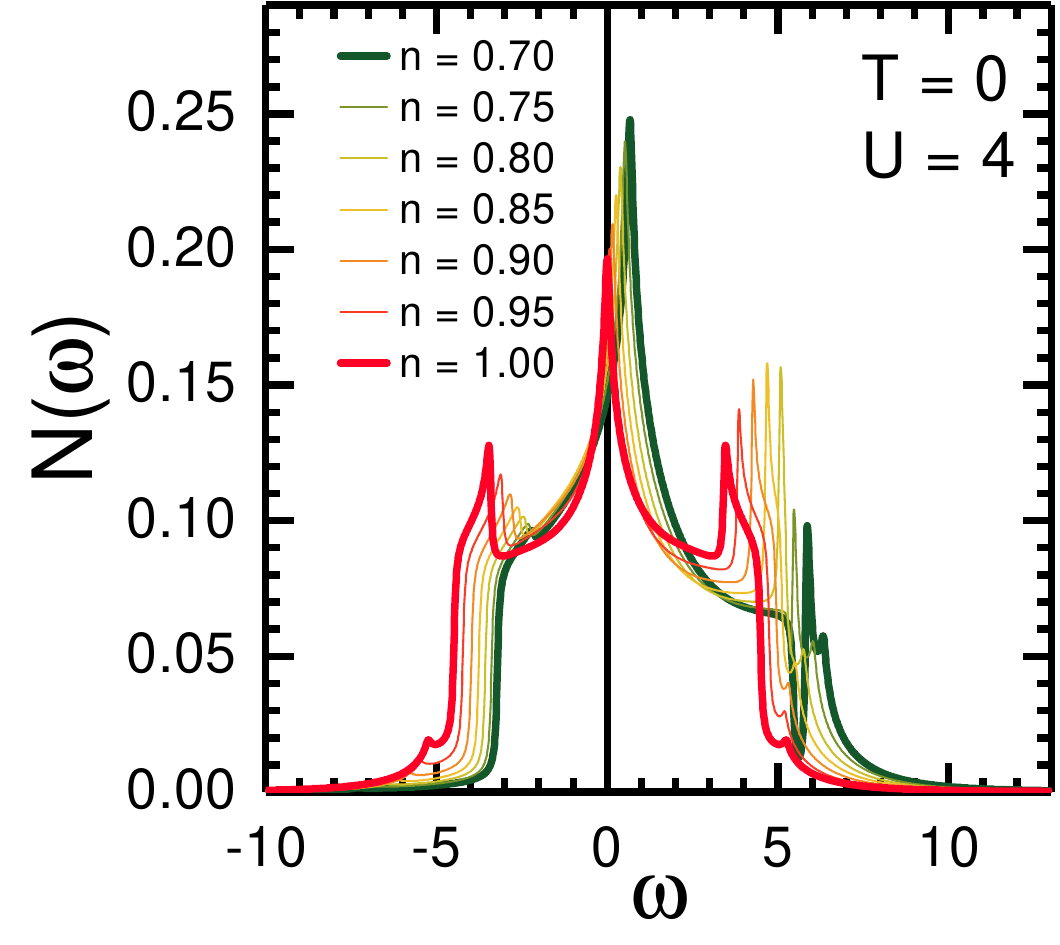} & \includegraphics[height=5.5cm]{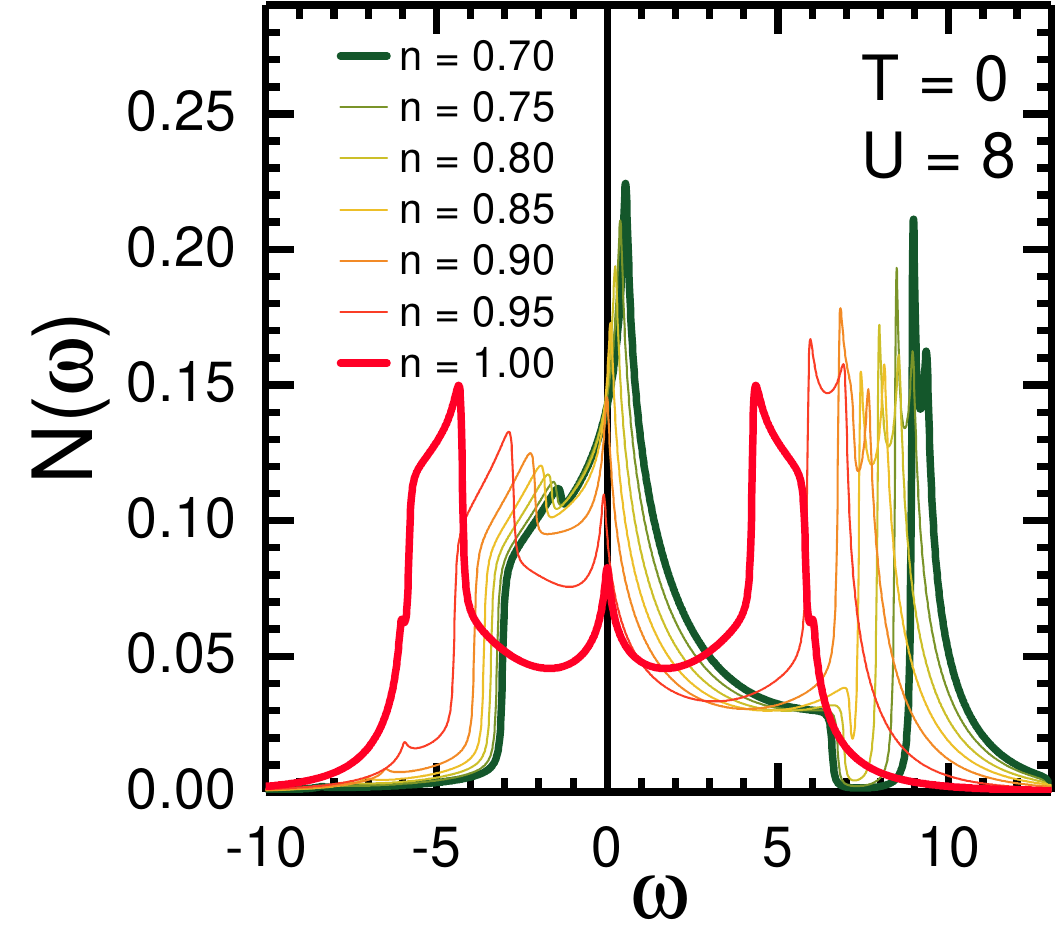}\tabularnewline
\end{tabular}
\par\end{centering}
\caption{(top row) Energy bands $E^{\left(\nu\right)}\left(\mathbf{k}\right)$
along the principal directions of the first Brillouin zone ($\Gamma=\left(0,0\right)\rightarrow S=\left(\pi/2,\pi/2\right)\rightarrow M=\left(\pi,\pi\right)\rightarrow X=\left(\pi,0\right)\rightarrow\Gamma=\left(0,0\right)$)
for $T=0$ and $U=4$ (left) and $U=8$ (right). The thickness of
each band is proportional to the value of the corresponding electronic
spectral density weight $\sigma_{cc}^{\left(\nu\right)}\left(\mathbf{k}\right)$.
The different colors correspond to different values of the doping
according to the legends in the bottom row. (bottom row) Density of
states $N\left(\omega\right)$ for $T=0$ and $U=4$ (left) and $U=8$
(right). The different colors correspond to different values of the
doping.\label{fig1}}
\end{figure}

In Fig.\ \ref{fig1}, the energy bands $E^{\left(\nu\right)}\left(\mathbf{k}\right)$
along the principal directions of the first Brillouin zone ($\Gamma=\left(0,0\right)\rightarrow S=\left(\pi/2,\pi/2\right)\rightarrow M=\left(\pi,\pi\right)\rightarrow X=\left(\pi,0\right)\rightarrow\Gamma=\left(0,0\right)$)
for $T=0$ and $U=4$ (top row, left panel) and $U=8$ (top row, right
panel) are reported. The thickness of each band is proportional to
the value of the corresponding electronic spectral density weight
$\sigma_{cc}^{\left(\nu\right)}\left(\mathbf{k}\right)$. The different
colors correspond to different values of the doping according to the
legends in the bottom row. At the smaller value of $U$, $U=4$, the
bands show a monotonous behavior on decreasing the doping down to
half filling, $n=1$, where the central band crosses the chemical
potential exactly along the main anti-diagonal ($X-S-Y=\left(0,\pi\right)$)
as in the non-interacting case. The great majority of the weight is
concentrated in the central band at all values of the filling and,
just close to half filling, the other two bands acquire some non negligible
weight for the momenta closer to the chemical potential. For the same
momenta, these two bands become almost completely flat. For the higher
value of $U$, $U=8$, the behavior is dramatically different at all
momenta and, in particular, at the $M$ point where the curves do
not follow anymore a monotonous behavior. The weights of the two non-central
bands result much higher, in particular, close to half filling, where
the first van Hove singularity (vHs) crossing seems to happen for
a value of doping smaller than $n=1$.

\begin{figure}[!p]
\noindent \begin{centering}
\begin{tabular}{cc}
\includegraphics[height=4cm]{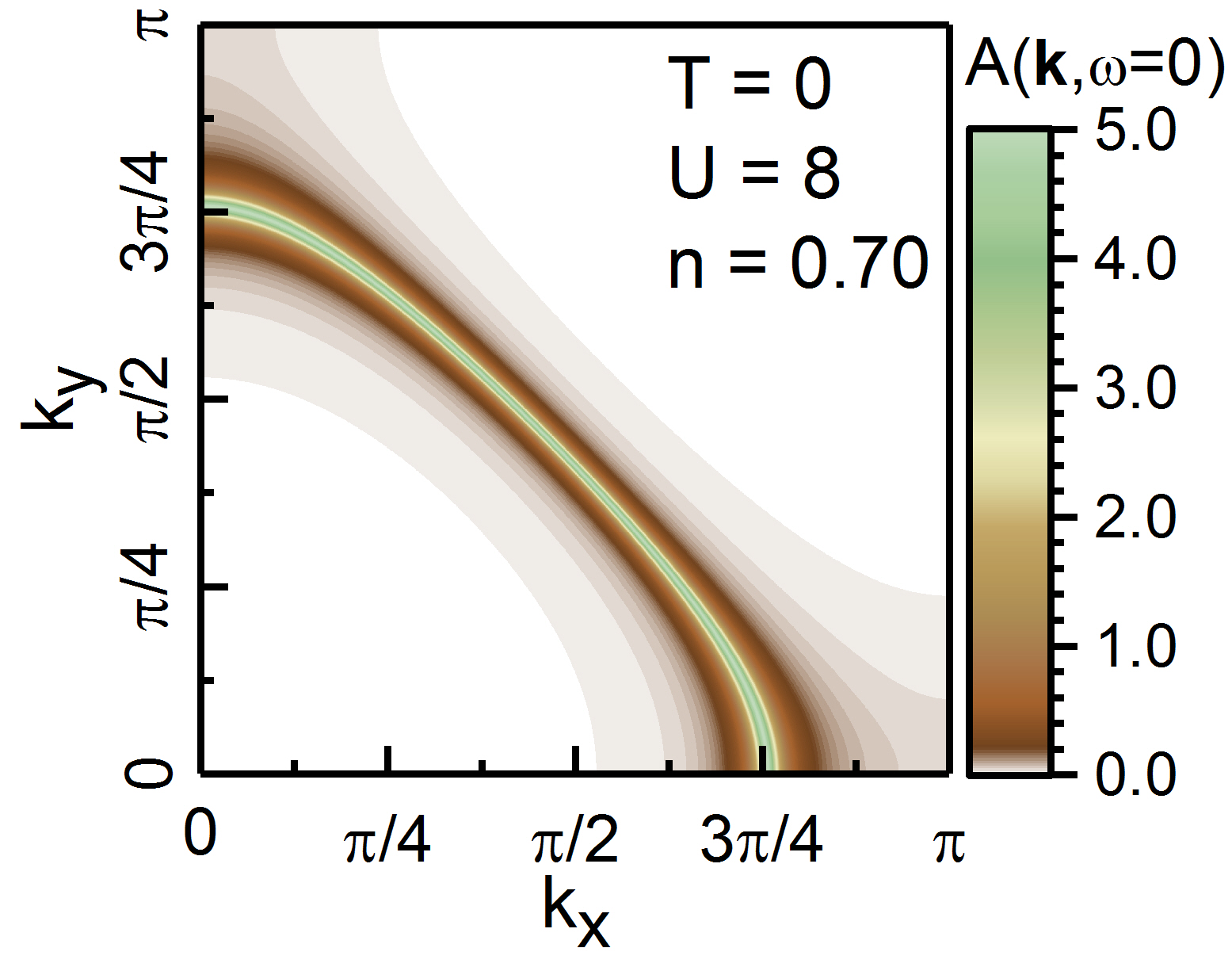} & \includegraphics[height=4cm]{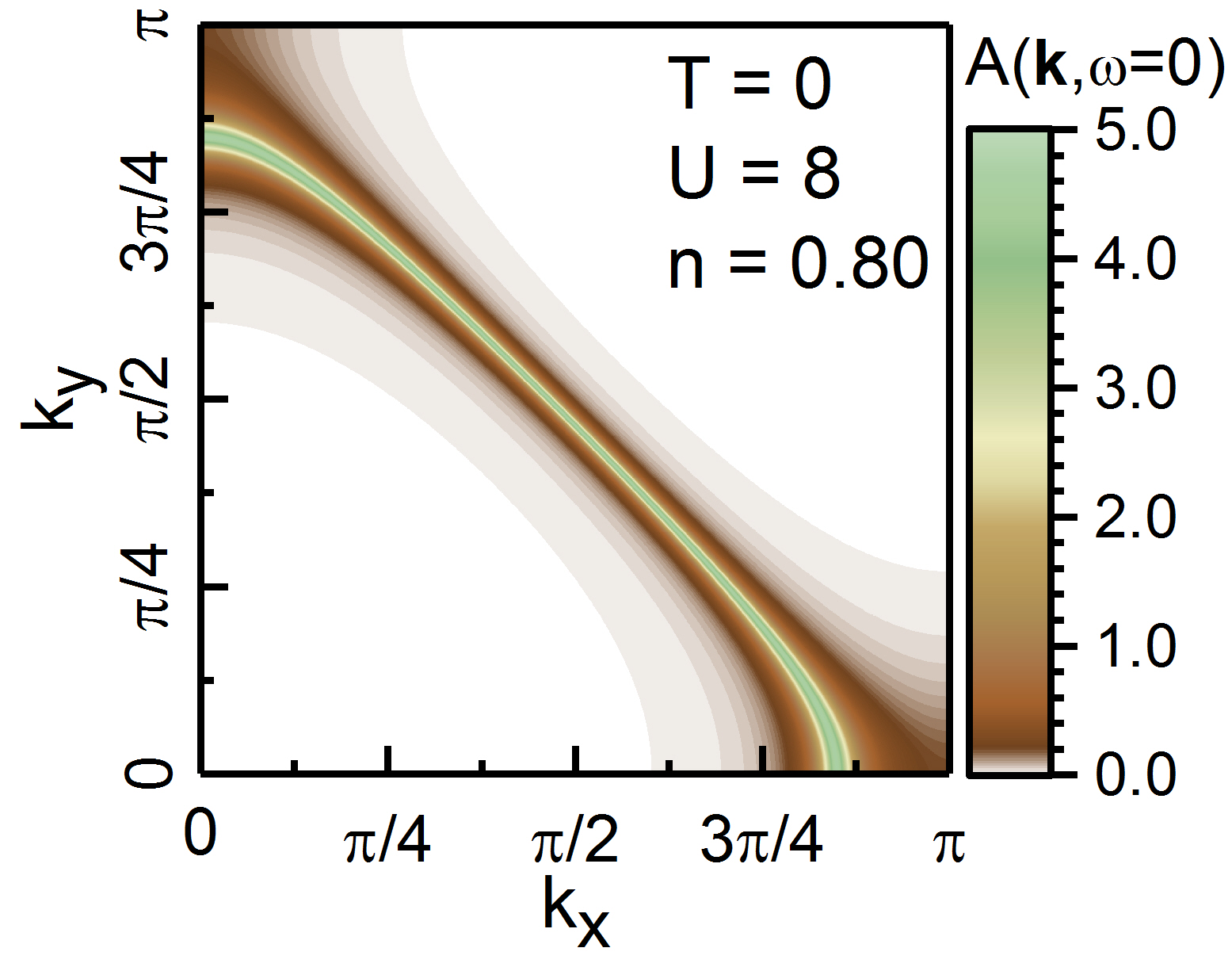}\tabularnewline
\includegraphics[height=4cm]{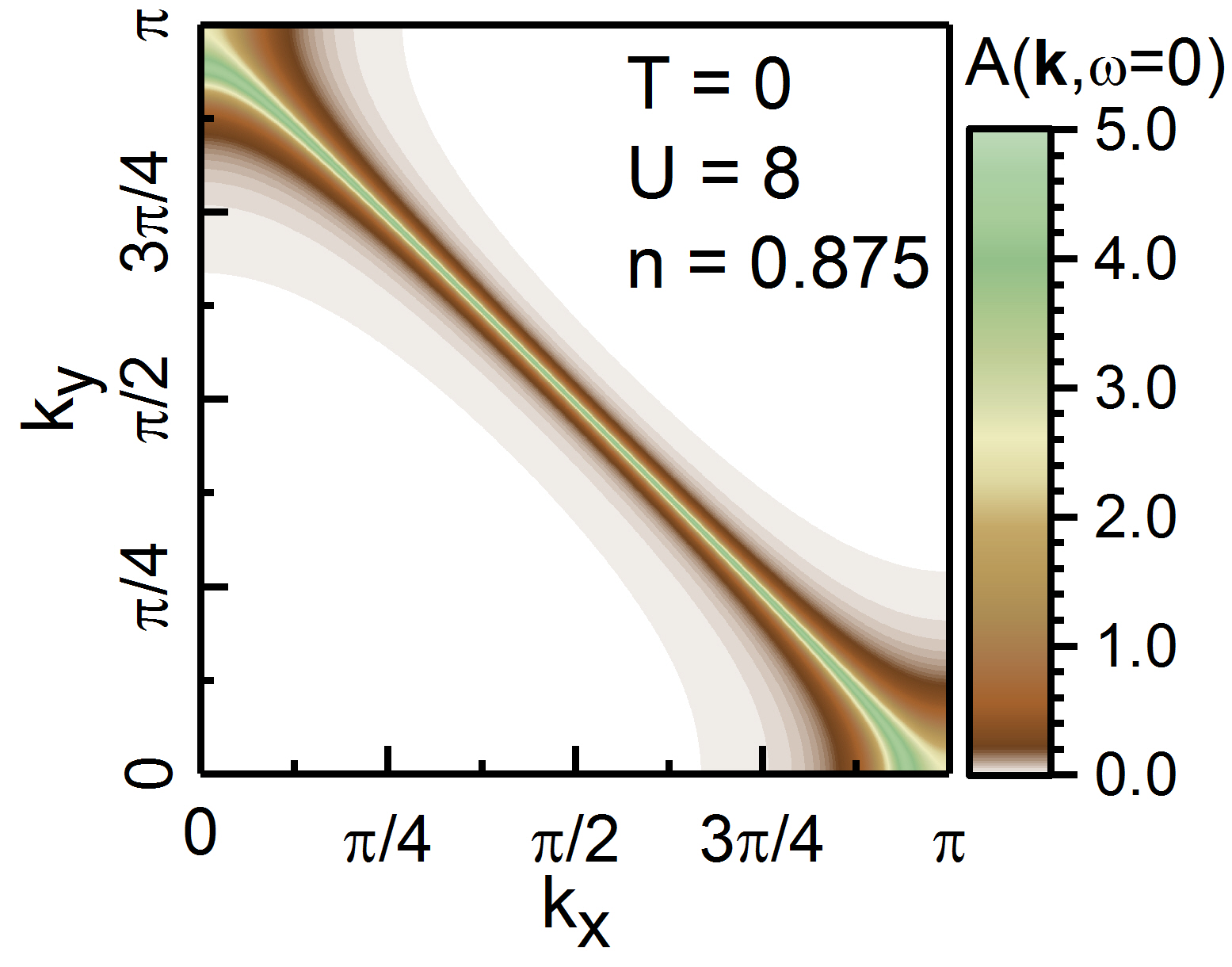} & \includegraphics[height=4cm]{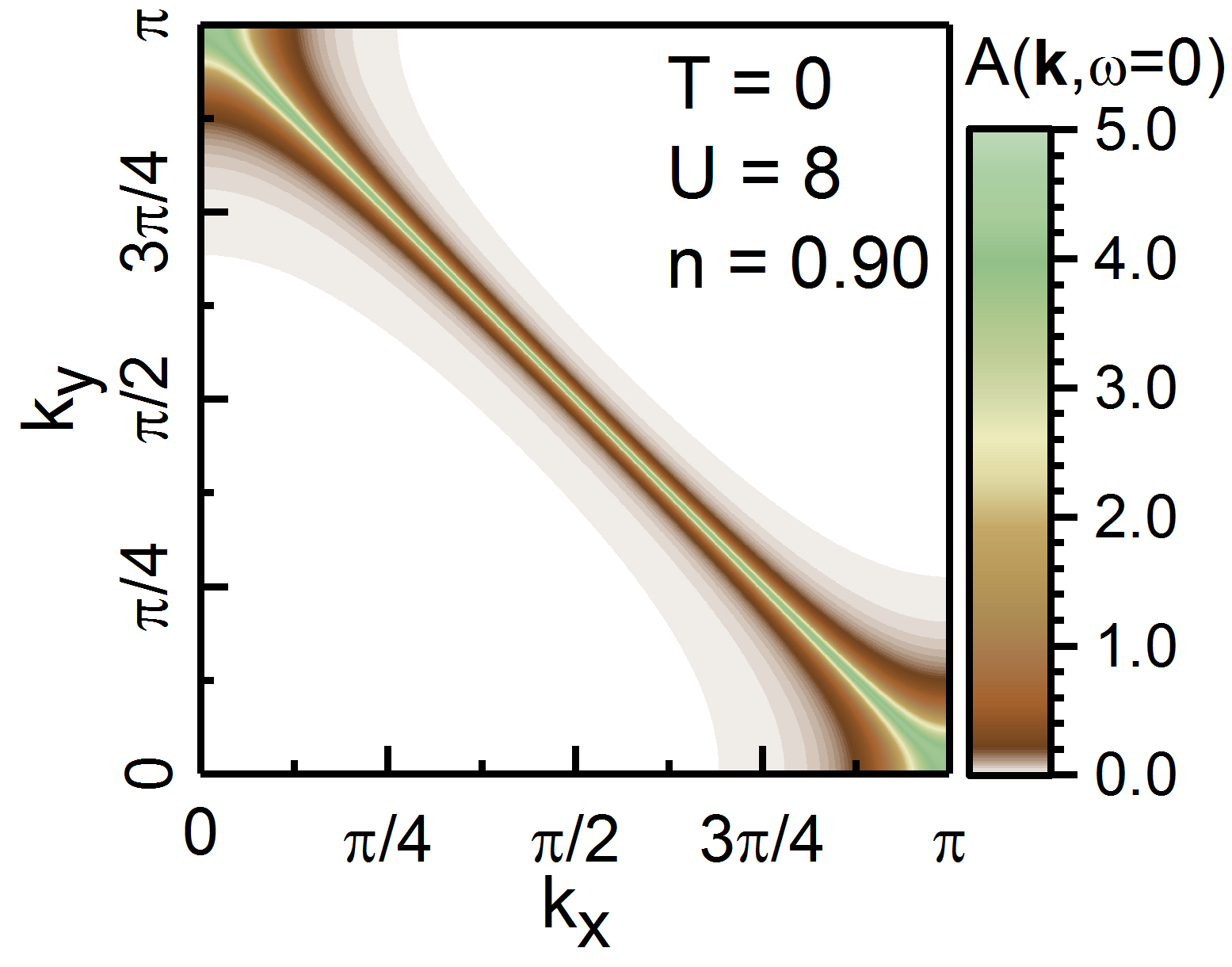}\tabularnewline
\includegraphics[height=4cm]{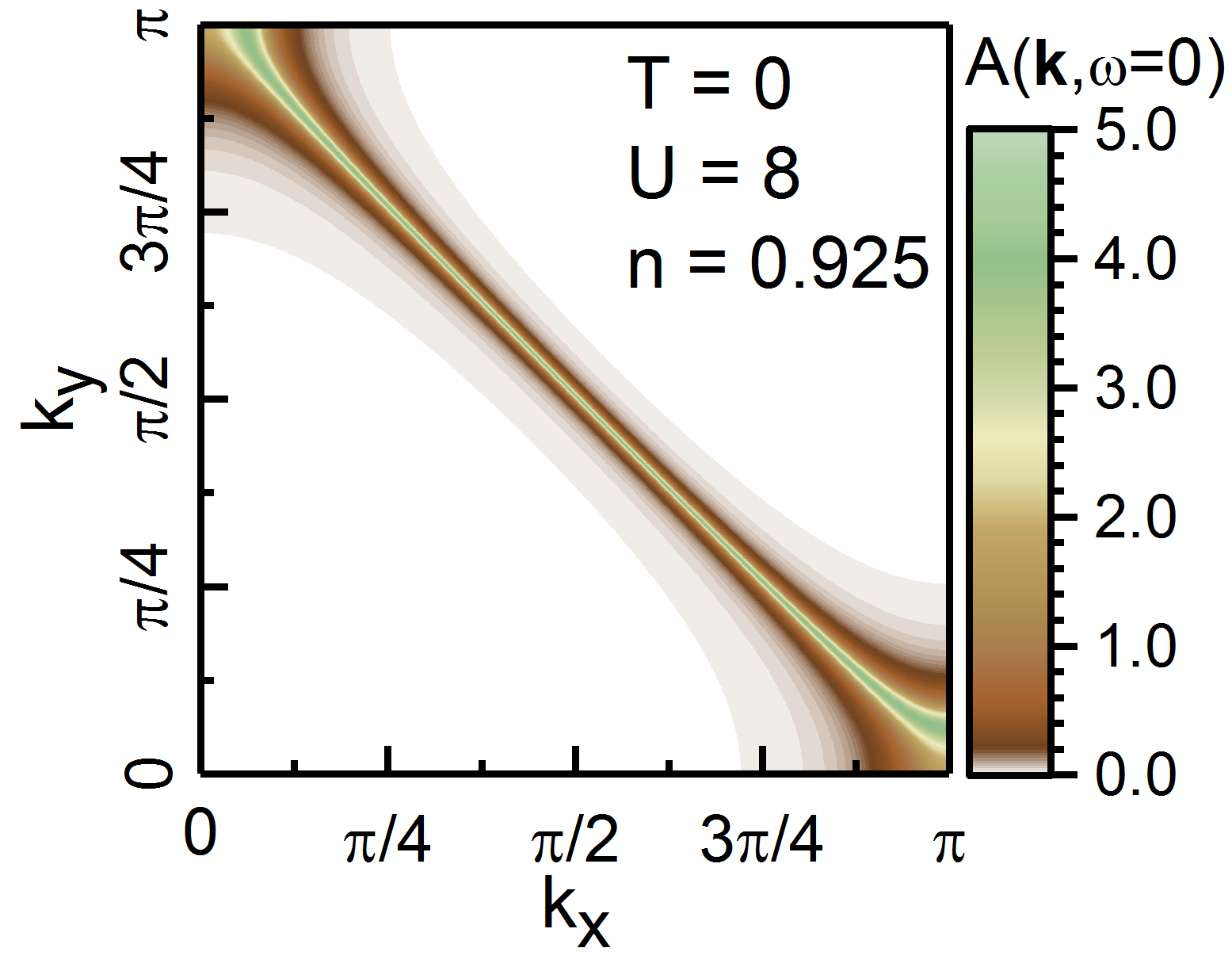} & \includegraphics[height=4cm]{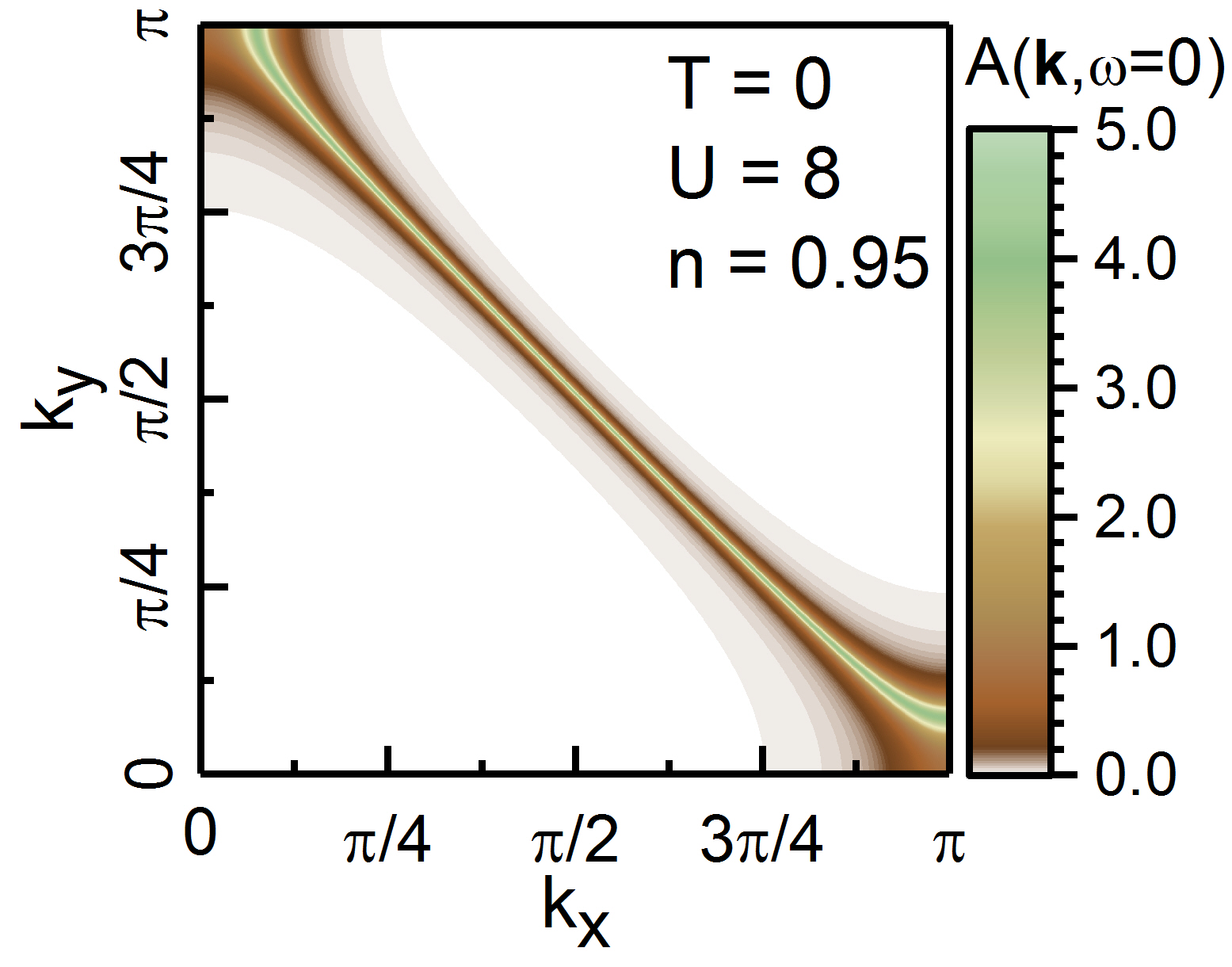}\tabularnewline
\includegraphics[height=4cm]{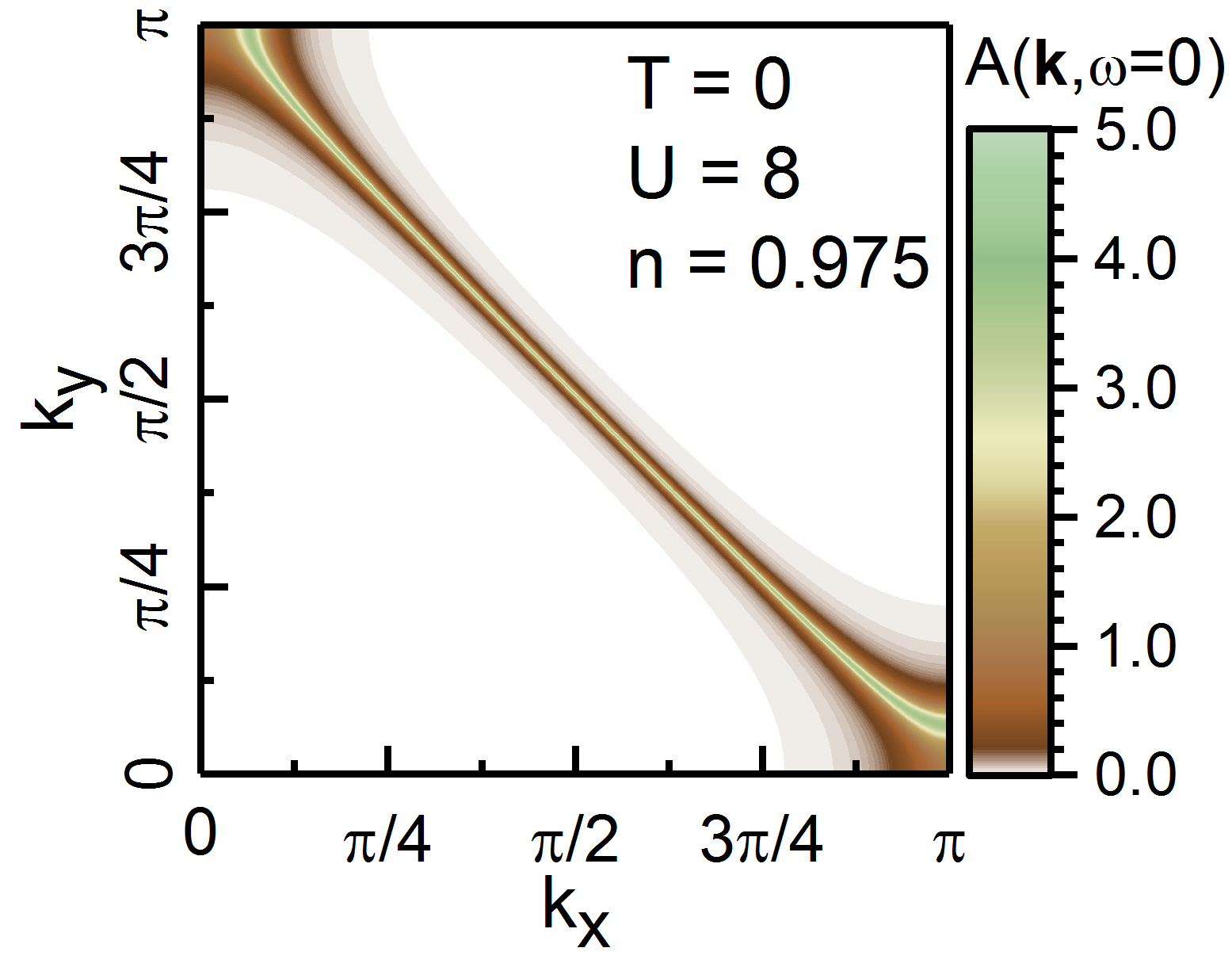} & \includegraphics[height=4cm]{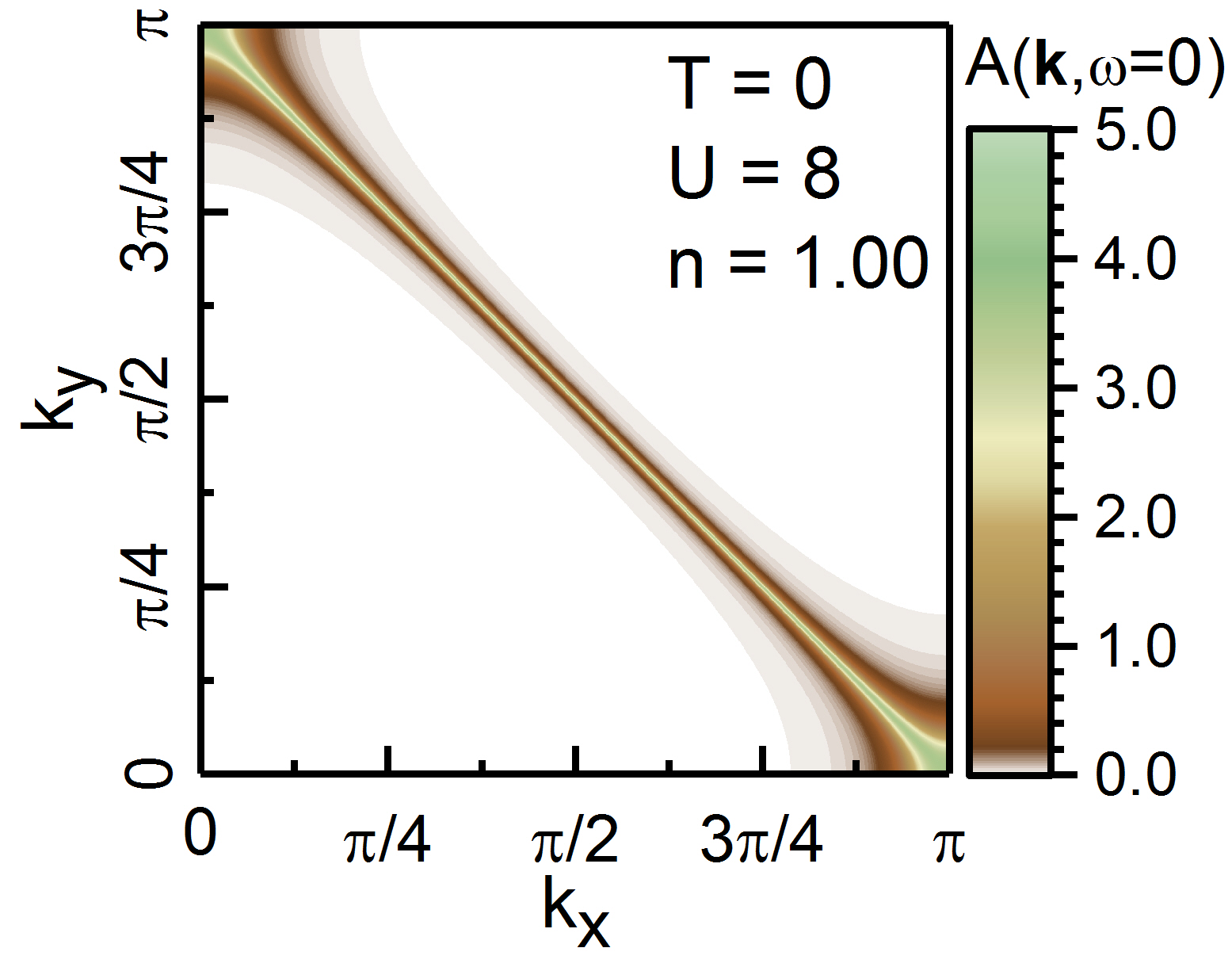}\tabularnewline
\end{tabular}
\par\end{centering}
\caption{Fermi surface in the top-right quadrant of the first Brillouin zone
read out through the maxima of $A\left(\mathbf{k},\omega=0\right)$
for various values of the doping at $T=0$ and $U=8$.\label{fig2}}
\end{figure}

In order to better explore the different features of the energy bands
and of the electronic weights, taking fully into account also the
actual shape/curvature of the energy bands, between the two values
of $U$, the density of states $N\left(\omega\right)$ for $T=0$
and $U=4$ (bottom row, left) and $U=8$ (bottom row, right) has also
been reported in Fig.\ \ref{fig1}. The legend clearly reports the
relationship between the different colors and the different values
of the doping analyzed. Again, for the smaller value of $U$, $U=4$,
the behavior is clearly monotonous and the vHs is clearly the higher
peak at all dopings and reach the chemical potential only once at
half filling following the evolution of the central band. The increase
of weight in the other two bands, together with their going flat close
to half filling, is now clearly visible in terms of well defined peak
structures surrounding the main central peak. The indications of an
unexpected and unconventional behavior coming from the energy bands
at the higher value of $U$, $U=8$, is completely confirmed by the
evolution of the density of states, which can also help us to better
understand which is the emergent behavior. First, it is now evident
that the vHs crossing happens twice: once at half filling as required
by the Luttinger theorem, but also at a lower filling ($n=n_{\mathrm{vHs}}\cong0.895$)
signaling the establishment of quite strong correlations modifying
the shape and the weights of the bands well beyond the weak-intermediate
coupling limit well represented by the $U=4$ results. Moreover, the
central peak is not anymore the highest one (at sufficiently low dopings):
the two facts together can be interpreted as the precursors of the
emergence of a pseudogap in the system in the region of doping close
to half filling. The peaks of the other two bands get evidently higher
and higher on decreasing the doping signaling the clear tendency towards
the opening of the Mott-Hubbard gap. This latter was always somehow
there between the two other bands, but the central band was filling
it in for lower values of $U$. This mechanism will lead to a finite
value of $U$ for the metal-insulator transition (MIT), that seems
already close for $U=8$.

In Fig.\ \ref{fig2}, we report the doping evolution of the Fermi
surface in the top-right quadrant of the first Brillouin zone read
out through the maxima of $A\left(\mathbf{k},\omega=0\right)$ at
$T=0$ and $U=8$. It is now clear that the Fermi surface changes
concavity not at half filling as for $U=4$, according to the conventional
weak-intermediate coupling scenario leading to a Fermi-like liquid
abiding the Luttinger theorem, but at $n\cong0.895$ leading to a
clear violation of the Luttinger sum rule. This can be explained only
taking into account that, in particular in the strong coupling regime,
the new quasi-particles establishing in the system, and replacing
the original electrons, are composite operators not satisfying canonical
commutation relations and, accordingly, whose Green's function is
not bound to obey the Luttinger theorem.

\begin{figure}[!t]
\noindent \begin{centering}
\begin{tabular}{cc}
\includegraphics[height=4.5cm]{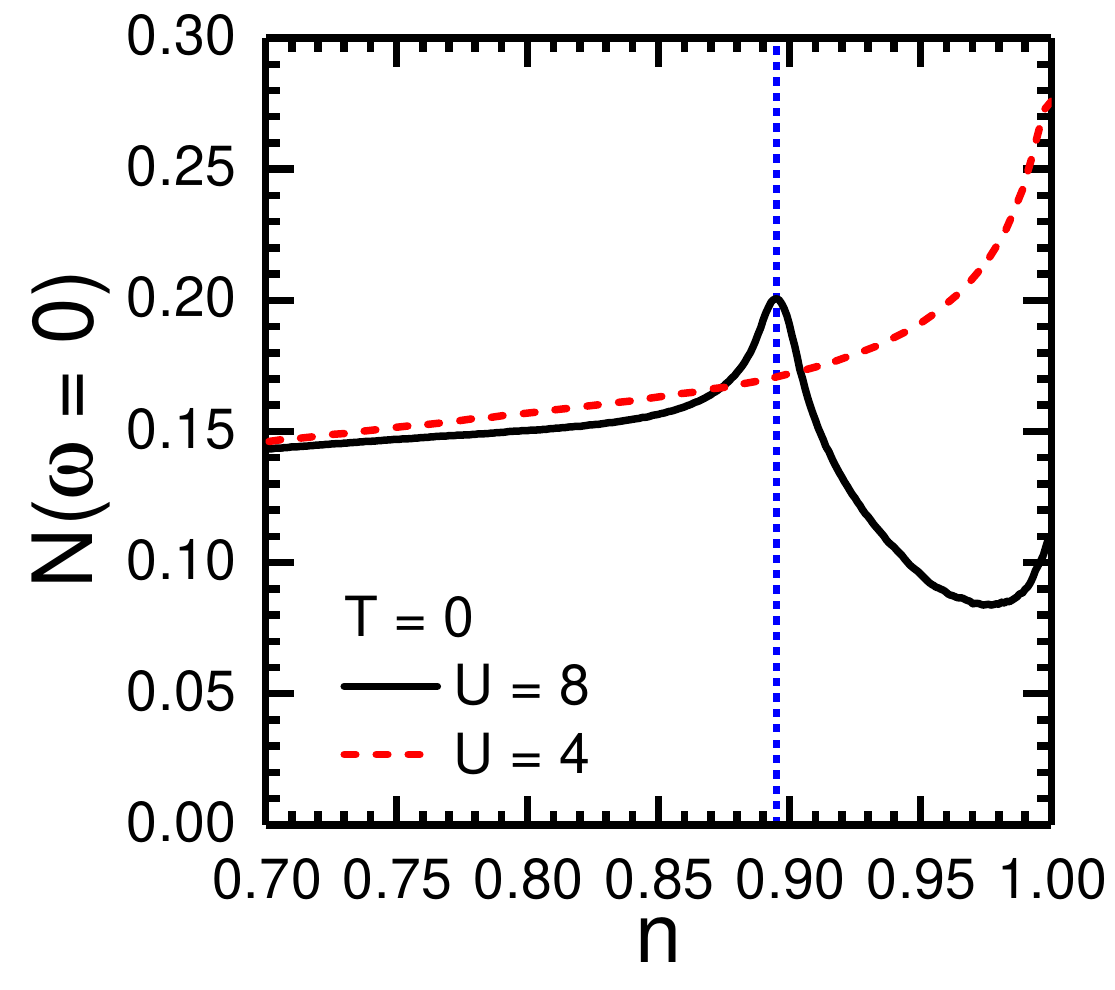} & \includegraphics[height=4.5cm]{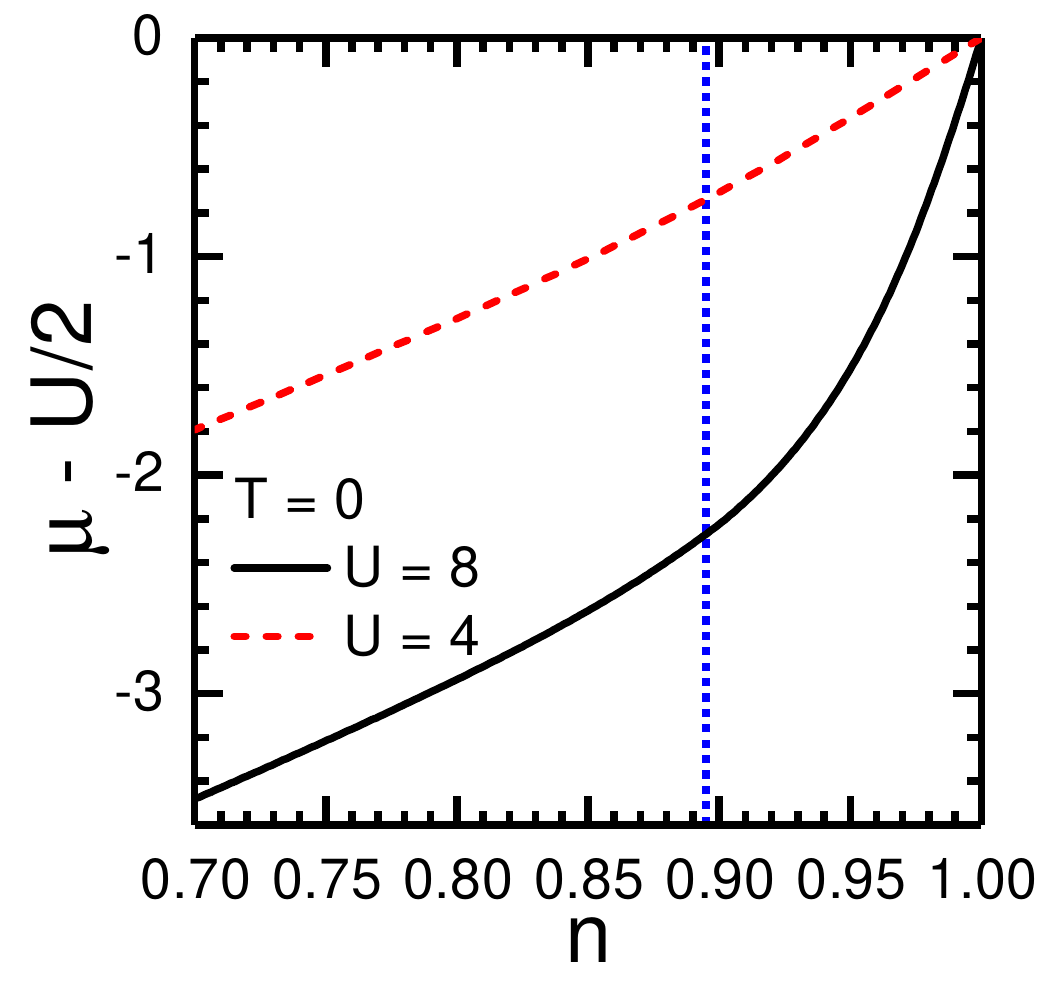}\tabularnewline
\includegraphics[height=4.5cm]{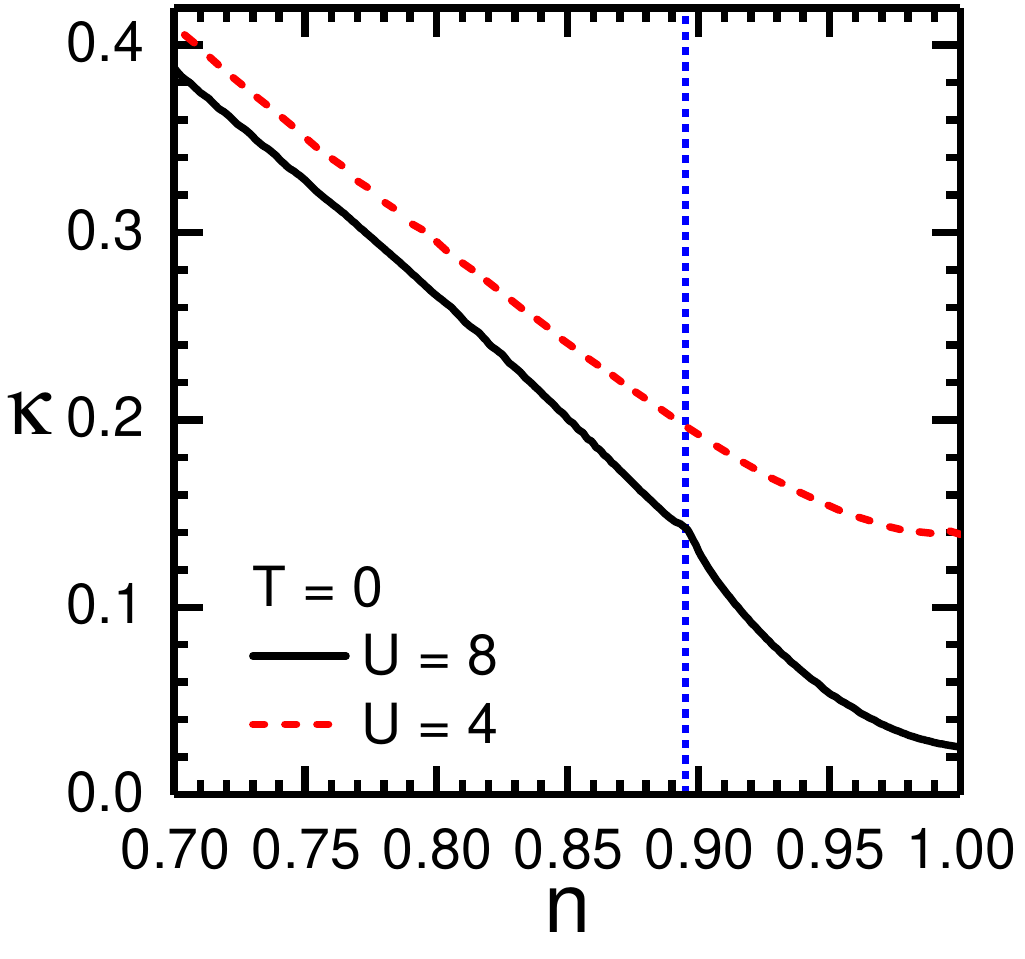} & \includegraphics[height=4.5cm]{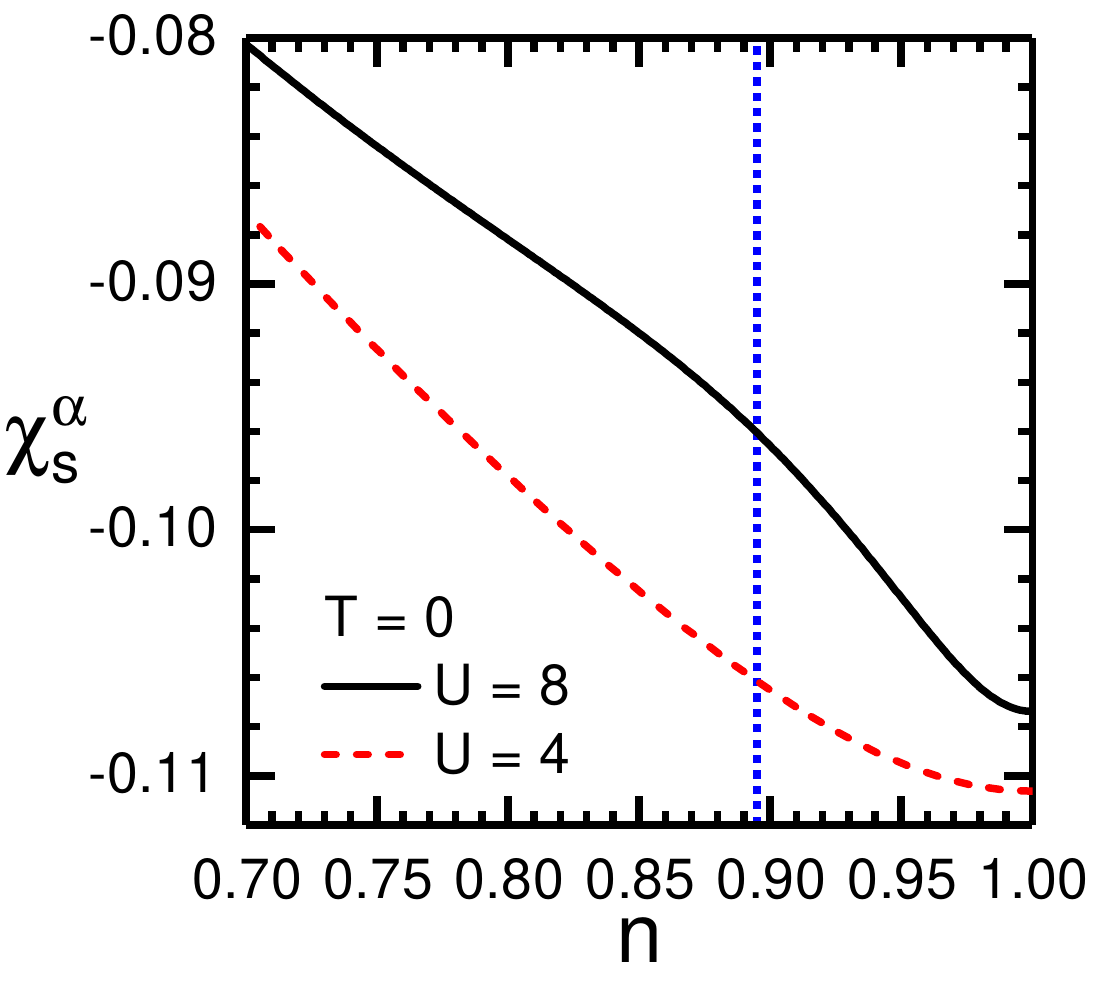}\tabularnewline
\end{tabular}
\par\end{centering}
\caption{Density of states at the chemical potential $N\left(\omega=0\right)$
(top row, left), chemical potential $\mu$ (top row, right), compressibility
$\kappa$ (bottom row, left) and nearest-neighbor spin-spin correlation
function $\chi_{s}^{\alpha}$ (bottom row, right) as functions of
the filling $n$ for $T=0$ and $U=4$ (red dashed line) and $U=8$
(black solid line). In all panels, the dotted blue line marks the
filling at which $N\left(\omega=0\right)$ has a maximum.\label{fig3}}
\end{figure}

Finally, in Fig.\ \ref{fig3}, we report the density of states at
the chemical potential $N\left(\omega=0\right)$ (top row, left),
the chemical potential $\mu$ (top row, right), the compressibility
$\kappa=\frac{1}{n^{2}}\frac{\partial n}{\partial\mu}$ (bottom row,
left) and the nearest-neighbor spin-spin correlation function $\chi_{s}^{\alpha}$
(bottom row, right) as functions of the filling $n$ for $T=0$ and
$U=4$ (red dashed line) and $U=8$ (black solid line). In all panels,
the dotted blue line marks the filling at which $N\left(\omega=0\right)$
has a maximum ($n=n_{\mathrm{vHs}}$). Looking at the doping evolution
of the density of states at the chemical potential $N\left(\omega=0\right)$,
it is now clear the fundamental difference of behavior between the
solution at $U=4$ and at $U=8$. The former shows the typical behavior
of a metal, the latter is on the verge of a MIT that is approached
in a very unusual way: the vHs is efficiently crossed, with the related
enhancement of the density of states, for a value of the filling smaller
than $n=1$, namely $n=n_{\mathrm{vHs}}$. This can be clearly seen
also in the chemical potential $\mu$ where the sudden change of slope
at the same value of filling is more than evident as well as the clear
tendency to reach a $\mu^{-}$ value, as it would happen at the MIT,
rather than $\frac{U}{2}$. Even more, the compressibility $\kappa$
reports a clear kink at the same value of filling and a sudden reduction
for smaller values of the doping, clearly signaling the upcoming emergence
of an instability driven by the proximity to the MIT. As a matter
of fact, it is the whole region of doping between $n=n_{\mathrm{vHs}}$
and half filling to exhibit clear fingerprints of strong correlations.
In fact, the nearest-neighbor spin-spin correlations, captured by
$\chi_{s}^{\alpha}$, show a net increase of intensity in that entire
region of doping, also denouncing the origin of such behavior. The
antiferromagnetic correlations, although still short range, appear
to be strong enough, in the proximity of an incipient MIT, to dramatically
modify the nature of the elementary excitations emerging in the system
leading to the necessity of a description in terms of composite operators
not necessarily obeying canonical commutation relations, whose treatment
definitely requires an operatorial approach.

\section{Conclusions}

We have studied the single-particle properties of the 2D Hubbard model
using the Composite Operator Method within a novel three-pole approximation
whose operatorial basis includes, together with the usual Hubbard
operator, a third field describing the electronic transitions dressed
by the nearest-neighbor spin fluctuations. These latter have proved
to play a crucial role in the unconventional behavior of the energy
bands, the density of states and the Fermi surface of the system in
the strong coupling regime ($U\geq8$). In particular, so strong,
although still short range, spin-spin correlations lead to the violation
of the Luttinger sum rule that can be seen as a precursor of a pseudogap
regime in proximity of an incipient MIT. The analysis of the doping
evolution of the density of states at the chemical potential, of the
chemical potential itself, of the compressibility and of the nearest-neighbor
spin-spin correlation function provides further evidence that this
scenario seems to be the one realized in the system. These findings
also prove the necessity of a description of so strongly correlated
systems in terms of composite operators, not necessarily obeying canonical
commutation relations, within an operatorial approach.



\end{document}